\documentclass[aps,pra,twocolumn,groupedaddress,showpacs,10pt]{revtex4-1}
\usepackage{hyperref}

\begin{document}


\title{Inferring the Gibbs state of a small quantum system}


\author{Jochen Rau}
\email[]{jochen.rau@q-info.org}
\homepage[]{www.q-info.org}
\affiliation{Institut f\"ur Theoretische Physik,
Johann Wolfgang Goethe-Universit\"at,
Max-von-Laue-Str. 1, 
60438 Frankfurt am Main,
Germany}


\date{\today}

\begin{abstract}
Gibbs states are familiar from statistical mechanics,
yet their use is not limited to that domain.
For instance,
they also feature  in the maximum entropy reconstruction of quantum states from incomplete measurement data.
Outside the macroscopic realm, however,
estimating a Gibbs state is a nontrivial inference task, due to two complicating factors:
the proper set of relevant observables might not be evident \textit{a priori;} 
and whenever data are gathered from
a small sample only, the best estimate for the Lagrange
parameters is invariably affected by the experimenter's
prior bias. 
I show how the two issues can be tackled with the help of Bayesian model selection and Bayesian interpolation, respectively,
and illustrate the use of these Bayesian techniques with a number of simple examples.
\end{abstract}


\maketitle

\section{\label{intro}Introduction}

Quantum states are not accessible to direct observation,
and so do not constitute \textit{per se} a physical reality.
Rather, they provide a convenient mathematical summary of an agent's expectations as to the outcomes of future experiments \cite{caves:quantumasbayes}.
Such expectations are formed both on the basis of past measurement data
and on the basis of any prior knowledge (say, about specific symmetries) that the agent may have.
In practice, the available experimental data are often far from perfect:
measurement devices work with limited accuracy;
sample sizes are finite;
and the set of observables measured might not be informationally complete.
Under such circumstances, 
a quantum state represents merely a \textit{model}, and hence a hypothesis, which is subject to testing, debate, and modification.
The more complex the physical system under study, and the sketchier the available data,
the more this model will be informed by the agent's prior knowledge.

Prior knowledge may be of two types:
(i)
the expectation, often based on symmetry considerations, that the quantum state has a certain {\textit{parametric form;}}
and 
(ii)
given a parametric form (including free-form as a special case), a bias as to its \textit{parameter values}.
Making proper use of such prior knowledge can lead to significant gains in the efficiency and accuracy of quantum-state tomography,
i.e., the reconstruction of a quantum state from imperfect data.
One recent example where prior knowledge about the parametric form has been exploited to great advantage, is the polynomial scheme for reconstructing near matrix product states \cite{cramer:efficient}.
The second type of prior knowledge, on the other hand, has been used in recent Bayesian modifications to the conventional maximum likelihood tomography scheme  \cite{audenaert:kalman,blume-kohout:hedged,rau:evidence}.

One parametric form that occupies a special place in physics is that of a \textit{Gibbs state}.
Such a state maximizes the entropy, or more generally,
minimizes the relative entropy with respect to some reference state, under given constraints on some selected set of expectation values.
Gibbs states are familiar from statistical mechanics where, 
in both the classical \cite{jaynes:info1,jaynes:info2} and the quantum  \cite{balian+balazs} case,
the principle of maximum entropy has long been recognised as the appropriate prescription for constructing the macrostate.
The common justification of this principle rests on a number of assumptions:
(i)
the system under consideration may be viewed as one constituent of a larger ensemble of identically prepared systems,
whose size approaches infinity (the ``thermodynamic limit'');
(ii)
pertaining to the global state of this fictitious infinite ensemble, there are constraints in the form of sharp values for the totals of certain  observables deemed relevant;
and
(iii)
there is clarity as to which observables are relevant.
While the first two assumptions are of a purely statistical nature,
the last implicitly invokes the system's dynamics. 
In equilibrium statistical mechanics, the relevant observables are the system's constants of the motion;
whereas in  nonequilibrium transport equations, they typically comprise the slowly varying degrees of freedom \cite{rau:physrep}.

Gibbs states play an important role even in realms where the above assumptions are not justified.
For instance, 
hadronization in $e^+e^-$ collisions is described with thermal distributions, even though the number of hadrons produced in one collision is hardly more than a handful \cite{becattini:thermal}.
Another example
is the extension of thermodynamics to nanoscale quantum systems,
and in particular, to the study of work extraction from such finite systems \cite{allahverdyan:nano,allahverdyan:work,janzing:molecular,dahlsten:workvalue}.
And finally,
in incomplete quantum-state tomography, Gibbs models feature in the reconstruction schemes  
based on  maximum entropy \cite{buzek:spin,buzek:aop,buzek:reconstruction,buzek:lnp},
or in case there is an initial bias towards some non-uniform reference state, on the principle of minimum relative entropy \cite{olivares+paris}.
In all these examples,
the measurement data, and hence any derived constraints, do not pertain to (quasi) infinite ensembles but to real samples which are small;
and
the systems are often too simple to exhibit a clear hierarchy of time scales that would lead to an obvious choice for the relevant observables.

When prior knowledge suggests that a small physical system is well described by a Gibbs state,
yet the proper set of relevant observables is not evident \textit{a priori},
the  choice of the latter becomes  a matter of statistical inference.
Competing theories might propose different sets of relevant observables;
and the task is then to decide rationally between them on the basis of rather sketchy data.
Typically, this inference task involves a trade-off between goodness-of-fit on the one hand,
favoring a large number of relevant observables;
and simplicity on the other (``Occam's razor''),
favoring a number that is as small as possible.
The appropriate framework for deciding such a trade-off is \textit{Bayesian model selection} \cite{sivia:modelselection}.
Adapting this framework to the task of finding the optimal set of relevant observables in a Gibbs model,
be it classical or quantum,
is one central objective of the present paper.

Given the set of relevant observables, the next inference task is the estimation of the associated Lagrange parameters.
Whenever the sample size is small, 
the estimate must take into account not only experimental data but also prior expectations.
The fact that the prior bias invariably exerts an influence on the parameter estimate,
is readily seen in a trivial example:
If five tosses of a coin yield ``heads'' five times, one is not yet ready to abandon one's prior bias towards a more or less fair coin;
only as evidence to the contrary accumulates, does this belief gradually erode.
The challenge, then, is to find the relative weights to be attributed to prior bias and data.
Again, the appropriate tools are furnished by Bayesian theory,
namely \textit{Bayesian interpolation} \cite{mackay:interpolation} in combination with the \textit{evidence procedure} \cite{rau:evidence}.
Putting these tools to use for the estimation of Lagrange parameters in a Gibbs model,
is the second main objective of the present paper.

The paper is organised as follows.
In Sec. \ref{significance}, I will start with some preliminaries about the $\chi^2$ distribution, the entropy concentration theorem, and the concept of statistical significance,
which will be needed in subsequent arguments.
Then I shall turn to the two inference problems outlined above, albeit in reverse order.
In Sec. \ref{estimation},
I will assume that the relevant observables of a Gibbs model are given, and show how its Lagrange parameters can be estimated in a way that accounts for both prior knowledge and measured data.
The estimation procedure will yield not just the optimal values for the Lagrange parameters,
but also the associated error bars.
In Sec. \ref{levels},
I shall consider the issue of the proper set of relevant observables, and show that Bayesian model selection provides a rational framework for choosing between rival proposals.
I will illustrate this method with two examples,
the classical analysis of Wolf's die (Sec. \ref{wolf}) and the quantum problem of deciding between an Ising and a Heisenberg description of an assembly of qubits (Sec. \ref{qubits}).
In Sec. \ref{conclusions}, I shall conclude with a brief summary.

There are a number of  appendices in which I collect
technical definitions and results that might not be familiar to readers of this journal,
yet whose inclusion in the main body of the text would render the flow of exposition unnecessarily cumbersome.
Specifically,
in Appendix \ref{lod},
I shall introduce the notion of a level of description;
in Appendix \ref{relpart},
the notions of coarse graining, relevant part of a state, and generalised Gibbs states;
and in Appendix \ref{geometry},
the definition and basic geometry of a Gibbs manifold,
which includes as a special case (discussed in Appendix \ref{bloch}) the geometry of the Bloch sphere.
In Appendix \ref{entropic},
I will introduce the concept of an entropic distribution on the Gibbs manifold;
and in Appendix \ref{approximations},
I will consider the meaning of the Gaussian approximation and of the thermodynamic limit.
Finally,
in Appendix \ref{statmech},
I shall connect the general framework of Gibbs models to the familiar terminology and basic relations of thermodynamics.

\section{\label{significance}Statistical significance}

In a generic experiment, some selected set of observables,
spanning the  \textit{{experimental level of description}} ${\cal F}$,
is measured
on $N$ identically prepared copies of a physical system,
yielding sample means $f$.
Given a  reference state $\sigma$,
these experimental data can be represented as 
a Gibbs model $\mu\in\pi^\sigma_{\cal F}({\cal S})$,
with Lagrange parameters adjusted such as to reproduce the observed sample means, $f(\mu)=f$.
The correspondence $f\leftrightarrow\mu$ is one-to-one.
On the same Gibbs manifold, let $\rho$ denote a theoretical model 
yielding expectation values $f(\rho)$;
these generally differ from the observed sample means.
As long as the difference is small,
the relative entropy between data and theoretical model
is approximately quadratic in the differentials $\delta f$.
According to Eq. (\ref{relentropy_nearby}), it is
\begin{equation}
	2N S(\mu\|\rho)
	\approx
	\chi^2(\mu\|\rho)
	,
\end{equation}
with 
\begin{equation}
	\chi^2(\mu\|\rho):=N \sum_{ab} (C^{-1})^{ab} \delta f_a \delta f_b
	.
\end{equation}
For definitions of further mathematical objects used here (${\cal F},\pi^\sigma_{\cal F}({\cal S}),C^{-1}$),
see Appendices \ref{lod} through \ref{geometry}.

Since the theoretical model may contain parameters that have been fitted to the data,
the differentials $\delta f$ might not be all independent;
the number $k$ of independent differentials is generally smaller than the dimension of the Gibbs manifold.
Given a (possibly fitted) theoretical model, 
the likelihood that the $k$ remaining independent degrees of freedom
yield some $\chi^2$ in the interval $[x,x+dx]$
is determined for large $N$ by
the probability density function
\begin{equation}
	\mbox{pdf}(x|k)=2^{-k/2} \Gamma(k/2)^{-1} x^{k/2-1} \exp(-x/2)
	,
\end{equation}
known as the $\chi^2$ distribution \cite{jaynes:book}.
In this distribution,
the exponential factor stems from the quantum Stein lemma (\ref{quantumstein});
the power factor  from the $k$-dimensional volume element;
and the numerical factors ensure proper normalisation.
The $\chi^2$ distribution is peaked at $\chi^2_{\rm max}=k-2$ (for $k>2$),
and has expectation value and variance
\begin{equation}
	\langle \chi^2 \rangle=k
	\ ,\ 
	\mbox{var}(\chi^2) = 2k
	,
\end{equation}
respectively.
For large arguments ($x\gg k\ln k$) it has an exponential tail independent of $k$,
\begin{equation}
	\mbox{pdf}(x|k)
	\sim
	\exp(-x/2) 
	.
\end{equation}

The above distribution of $\chi^2$, and hence of relative entropy, implies the {\textit{entropy concentration theorem} \cite{jaynes:concentration}:}
As the sample size $N$ increases,
the  relative entropy between data and theoretical model is predicted to become more and more concentrated (with a width of order $1/N$)
around a smaller and smaller expectation value (also of  order $1/N$).
Relative entropy being approximately quadratic in the coordinate differentials $\delta f$, 
this implies as an immediate corollary that deviations between measured sample means and theoretical expectation values are expected to scale as $O(1/\sqrt{N})$.
The entropy concentration theorem can thus be employed to assess quickly the statistical significance of experimental deviations from theoretical predictions:
As long as the relative entropy between data and theoretical model is of the order $1/N$,  deviations likely fall within the range of statistical fluctuations;
yet as soon as their relative entropy exceeds this limit, deviations become significant and may indicate the need to revise the theoretical model.
This simple entropy test is closely related to the $\chi^2$ test in conventional statistics.
Theoretical models are typically rejected whenever at the observed $\chi^2$ the cumulative distribution function exceeds some predefined bound,  whose value in turn depends on the confidence level required.
The $\chi^2$ test then points to the need to revise a given theoretical model,
and may trigger creative thinking about possible alternatives.

\section{\label{estimation}Estimating Lagrange parameters}

I consider the situation where
it is assumed from the outset that 
a physical system is to be described by some Gibbs model $\omega\in\pi^\sigma_{\cal G}({\cal S})$,
with given reference state $\sigma$ and level of description ${\cal G}$,
yet unknown parameter values.
The initial uncertainty about the parameter values is reflected in a prior probability distribution 
$\mbox{prob}(\omega|\sigma,{\cal G})$ over the Gibbs manifold.
Subsequently, 
on a sample of size $N$,
one measures some set of sample means $f$.
The associated {{experimental level of description}} ${\cal F}$
may or may not coincide with the theoretical level of description ${\cal G}$.
In the light of the observed sample means and the prior distribution over the Gibbs manifold,
one wants to infer the most plausible estimate for $\omega$.

After collecting the experimental data,
the probability distribution over the Gibbs manifold must be updated 
according to the \textit{Bayes rule} \cite{schack:bayesrule}
\begin{equation}
	\mbox{prob}(\omega|f,N,{\cal F};\sigma,{\cal G})
	\propto
	\mbox{prob}(f|N,\omega,{\cal F})\, \mbox{prob}(\omega|\sigma,{\cal G})
	.
\label{bayesrule}
\end{equation}
The first factor on the right hand side is
the  \textit{likelihood} of observing the sample means $f$, given $\omega$;
it is 
\begin{equation}
	\mbox{prob}(f|N,\omega,{\cal F})
	\propto
	\int_{{\cal S}|_f}d\rho\,\mbox{prob}(\rho|N,\omega)
	,
\end{equation}
with the integration ranging over the submanifold ${\cal S}|_f$ of all states that satisfy the constraints $f(\rho)=f$,
and normalised according to Eq. (\ref{volume}), 
\begin{equation}
		\int \prod_b d f_b \sqrt{\det C^{-1}}	\mbox{prob}(f|N,\omega,{\cal F}) 
		= 
		1
		.
\end{equation}
For large $N$, 
by virtue of the quantum Stein lemma (\ref{quantumstein}) and the law of Pythagoras (\ref{pythagoras}),
the likelihood can be written as
\begin{equation}
	\mbox{prob}(f|N,\omega,{\cal F})
	\propto
	\exp[-NS(\mu\|\omega)]
	,
\label{likelihood}
\end{equation}
where $\mu\in\pi^\omega_{\cal F}({\cal S})$ is the unique Gibbs model associated with the measured $f$ and reference state $\omega$.
In other words, on the Gibbs manifold $\pi^\omega_{\cal F}({\cal S})$, the Gibbs model $\mu$ representing experimental data 
is distributed entropically around $\omega$, $\mu\sim\mbox{Ent}(N,\omega,{\cal F})$ (see Appendix \ref{entropic}).

The second factor on the right hand side of the Bayes rule (\ref{bayesrule}) is the \textit{prior}.
In principle, it can take any form;
there is no constraint as to the prior knowledge that an agent may have.
But there are good reasons to assume that it is entropic, too,
$\omega\sim\mbox{Ent}(\alpha,\sigma,{\cal G})$,
with its peak at some initial bias $\sigma$,
and the parameter $\alpha$ characterising the agent's degree of confidence as to this initial bias.
Conceptually, 
if the only prior knowledge available is the initial bias $\sigma$,
then one would demand of a prior distribution on $\pi^\sigma_{\cal G}({\cal S})$ that
it be peaked at and symmetric around this bias;
that it be form-invariant under coarse graining;
and that upon composition of systems,
it be non-committal as to any correlations between the systems.
As I discuss in more detail in Appendix \ref{entropic},
these requirements are satisfied by entropic distributions.
In addition,
an entropic prior is particularly convenient because 
it is (approximately) conjugate to the likelihood (\ref{likelihood}):
Upon any measurement that is informationally complete with respect to the unknown model parameters, 
${\cal F}\supset{\cal G}$,
Bayesian updating yields a posterior which (in the Gaussian approximation) is again entropic, 
and which differs from the prior only by a change of parameters, $(\alpha,\sigma)\to(\alpha',\sigma')$.

Assuming an entropic prior and making the Gaussian approximation,
the Bayes rule yields for ${\cal F}\supset{\cal G}$ the posterior (\ref{post_supset}),
and hence
\begin{equation}
	\mbox{prob}(\omega|f,N,{\cal F};\sigma,{\cal G})
	\propto
	\mbox{prob}(\omega|\alpha+N,\rho ,{\cal G})
	;
\end{equation}
whereas for ${\cal F}\subset{\cal G}$,
it yields the posterior (\ref{post_subset}),
and hence
\begin{eqnarray}
	\mbox{prob}(\omega|f,N,{\cal F};\sigma,{\cal G})
	&\propto&
	\mbox{prob}(\pi^{\sigma}_{\cal F}(\omega)|\alpha+N,\rho ,{\cal F})
	\times
	\nonumber \\
	&&
	\mbox{prob}(\pi^{\rho}_{\neg_{{\cal G},\rho}{\cal F}}(\omega)|\alpha,\rho,{\neg_{{\cal G},\rho}{\cal F}})
	.
\end{eqnarray}
In both cases,
the posterior is peaked at the model
\begin{equation}
	\rho \propto \exp\left[ 
	\frac{\alpha}{\alpha+N} \ln \sigma + \frac{N}{\alpha+N} \ln \pi^\sigma_{{\cal F}\cap{\cal G}}(\mu)
	\right]
	.
\label{post_estimate}
\end{equation}
This model constitutes the most plausible posterior estimate for $\omega$.

The posterior estimate for $\omega$ {interpolates} 
between initial bias and data, 
and depending on the relative sizes of $\alpha$ and $N$,
may attribute more weight to one or the other;
this is an example of \textit{Bayesian interpolation} \cite{mackay:interpolation}.
In the extreme case where the prior is sharply peaked while sample sizes are small, $N\ll\alpha$,
parameter estimation will be dominated by the prior, and one is therefore advised to stick to the initial bias, $\rho\approx\sigma$;
while in the opposite case where the prior is broad and sample sizes are big, $N\gg\alpha$,
parameter estimation will be dominated by the likelihood function,
and the best estimate for the model is close to the \textit{maximum likelihood} estimate 
$\rho\approx\pi^\sigma_{{\cal F}\cap{\cal G}}(\mu)$.
Attached to the estimate are error bars of the order $O(1/\sqrt{\alpha+N})$ as to those model parameters that have  been measured,
and in case ${\cal F}\subset{\cal G}$,
$O(1/\sqrt{\alpha})$ as to those that have not.

The above estimation procedure preserves the Gibbs form, in the following sense.
Whenever the prior bias $\sigma$ is a generalised Gibbs state,
with some level of description that encloses ${\cal F}\cap{\cal G}$,
the posterior estimate will retain this form,
\begin{equation}
	\sigma\in\pi^\tau_{\cal H}({\cal S})
	\ ,\ 
	{\cal H} \supset {\cal F}\cap{\cal G}
	\Rightarrow
	\rho\in\pi^\tau_{\cal H}({\cal S})
	,
\end{equation}
for arbitrary values of $\alpha$ and $N$.
In particular, 
it is $\rho \in\pi^\sigma_{{\cal F}\cap{\cal G}}({\cal S})$.
If on the Gibbs manifold $\pi^\tau_{\cal H}({\cal S})$ the prior bias has 
Lagrange parameters $\lambda(\sigma)$,
and the maximum likelihood estimate has Lagrange parameters $\lambda(\pi^\sigma_{{\cal F}\cap{\cal G}}(\mu))$,
then the Lagrange parameters of the posterior estimate (\ref{post_estimate}) are given by linear interpolation,
\begin{equation}
	\lambda(\rho)
	=
	\frac{\alpha}{\alpha+N} \lambda(\sigma)
	+
	\frac{N}{\alpha+N} \lambda(\pi^\sigma_{{\cal F}\cap{\cal G}}(\mu))
	.
\label{interpolate_lagrange}
\end{equation}

The posterior estimate depends critically on the parameter $\alpha$;
this parameter has so far been left unspecified.
Provided the experiment reveals a significant deviation from the initial bias 
(only then does the need arise to update this bias)
and is sufficiently detailed,
\begin{equation}
	\chi^2(\pi^\sigma_{\cal F}(\mu)\|\sigma) > \dim \pi^\sigma_{\cal F}({\cal S}) \gg 1
	,
\label{evidence_applicability}
\end{equation}
the optimal value for $\alpha$ can be estimated \textit{a posteriori} with the help of the {evidence procedure} \cite{rau:evidence}.
This procedure yields an interpolation parameter
\begin{equation}
	\alpha/(\alpha+N)
	\approx
	\dim \pi^\sigma_{\cal F}({\cal S}) / \chi^2(\pi^\sigma_{\cal F}(\mu)\|\sigma)
	.
\label{optimal_alpha}
\end{equation}
The estimate depends only on the experimental level of description ${\cal F}$,
but not on the theoretical level ${\cal G}$ employed for the Gibbs model $\omega$.

To illustrate the above framework, I consider the following simple example.
A source emits a physical system (say, a molecule) which can be in its ground state or in one of $24$ excited states;
this spectrum may or may not be degenerate.
Prior theoretical considerations suggest that the source is thermal, and hence,
that the occupation probabilities $\{p_i\}$, $i=0\ldots 24$, of the energy levels follow a canonical distribution.
There is uncertainty about the temperature, but according to initial estimates, it is expected to be around $100$K.
As regards the system's state, therefore, the initial bias is a canonical state $\sigma\propto\exp(-\beta H)$,
with level of description ${\cal H}=\mbox{span}\{1,H\}$,
uniform reference state,
and Lagrange parameter $\beta(\sigma)\approx 1/100 \mbox{K}$.
Then one performs $N=12,000$ runs of the experiment, and in each run, measures the actual occupation of the energy levels.
One finds that the measured distribution of relative frequencies $\{f_i\}$ differs from the expected $\{p_i\}$.
The observed mean energy, 
$\sum_i f_i E_i$,
corresponds to a temperature $110$K rather than $100$K;
and in addition, the shape of the observed distribution may or may not deviate from the canonical form.
Altogether, one finds that the data differ from prior expectation by a distance, say,
\begin{equation}
	\chi^2(\pi^\sigma_{\cal F}(\mu)\|\sigma) \approx 2N \sum_{i=0}^{24} f_i \ln (f_i/p_i) \approx 96
	.
\end{equation}
This deviation is significant enough, and the number of independent sample means 
($\dim \pi^\sigma_{\cal F}({\cal S})=24$) is sufficiently large, 
to satisfy both conditions in Eq. (\ref{evidence_applicability}).
So it is justified to apply the evidence procedure, yielding the interpolation parameter $\alpha/(\alpha+N)\approx 1/4$.
If one insists that the system be modelled by a canonical distribution, 
${\cal G}={\cal H}$,
then the posterior estimate for its inverse temperature is neither the initial $1/100$K nor the observed $1/110$K,
but the interpolation (\ref{interpolate_lagrange}),
which in this example yields $\beta(\rho)\approx 1/107.3 \mbox{K}$.

\section{\label{levels}Comparing levels of description}

Up to this point the level of description of the theoretical model, and hence the Gibbs manifold $\pi^\sigma_{\cal G}({\cal S})$ from which a model was to be selected, 
have been assumed to be given \textit{a priori.}
Now they will become themselves  subject to statistical inference.

If a model is to have any explanatory value, its number of  parameters must be strictly smaller than the number of data points;
and so its level of description must be a proper subspace of the space spanned by the measured observables, ${\cal G}\subset{\cal F}$.
In fact, in the spirit of Occam's razor one would always prefer simpler models over more complicated ones;
yet when this is taken too far, the fit with the data might deteriorate. 
Striking the right balance between simplicity and goodness-of-fit, and
determining thus the optimal level of description, constitutes  a non-trivial inference task.
In this Section, I shall discuss how the Bayesian framework for {model selection} can guide the proper choice of the level of description.
If presented with two rival proposals for the level of description, 
this framework allows one to evaluate their relative degree of plausibility in the light of experimental data and prior expectations.

If a $\chi^2$ analysis has revealed that observed deviations from model predictions are statistically significant,
one might consider moving to a more accurate model by expanding the level of description,
${\cal G} \to {\cal H}$, with ${\cal G}\subset{\cal H}\subset{\cal F}$.
Provided the  priors on the respective Gibbs manifolds $\pi^\sigma_{\cal G}({\cal S})$ and $\pi^\sigma_{\cal H}({\cal S})$ are both entropic around the same initial bias $\sigma$,
the relative plausibility of the two levels of description is given by the {Bayes rule},
\begin{equation}
	\frac{\mbox{prob}({\cal G}|\mu,N,{\cal F};\alpha,\sigma)}{\mbox{prob}({\cal H}|\mu,N,{\cal F};\alpha,\sigma)}
	=
	\frac{\mbox{prob}({\cal G})}{\mbox{prob}({\cal H})}
	\frac{\mbox{prob}(\mu|N,{\cal F};\alpha,\sigma,{\cal G})}{\mbox{prob}(\mu|N,{\cal F};\alpha,\sigma,{\cal H})}
	.
\end{equation}
Here ${\cal F}$ denotes the experimental level of description,
$N$ the sample size, and $\mu\in\pi^\sigma_{\cal F}({\cal S})$ the Gibbs model associated with the measured data.
The parameter $\alpha$, which characterises the degree of confidence as to the initial bias, is assumed to be identical for both entropic priors;
this assumption is corroborated by the estimate (\ref{optimal_alpha}),
which does not depend on the level of description employed for a theoretical model.

The first factor on the right hand side is the ratio of prior preferences which, to be fair, is often taken to be of order $1$.
The second factor can be calculated via marginalisation,
\begin{eqnarray}
	&&
	\mbox{prob}(\mu|N,{\cal F};\alpha,\sigma,{\cal G})
	=
	\nonumber \\
	&&
	\int_{\pi^\sigma_{\cal G}({\cal S})} d\omega\,
	\mbox{prob}(\mu|N,\omega,{\cal F})\,
	\mbox{prob}(\omega|\alpha,\sigma,{\cal G})
	,
\end{eqnarray}
and likewise for ${\cal H}$.
In the Gaussian approximation,
the integrand is given by Eq. (\ref{post_supset});
which in the regime $N\gg\alpha$, with $\rho\approx\pi^\sigma_{\cal G}(\mu)$,
yields 
\begin{eqnarray}
	&&
	\mbox{prob}(\mu|N,{\cal F};\alpha,\sigma,{\cal G})
	\approx
	\nonumber \\
	&&
	\mbox{prob}(\mu|N,\pi^\sigma_{\cal G}(\mu),\neg_{{\cal F},\pi^\sigma_{\cal G}(\mu)}{\cal G})\,
	\mbox{prob}(\pi^\sigma_{\cal G}(\mu)|\alpha,\sigma,{\cal G})
	.
\end{eqnarray}
The ratio is then
\begin{equation}
	\frac{\mbox{prob}(\mu|N,{\cal F};\alpha,\sigma,{\cal G})}{\mbox{prob}(\mu|N,{\cal F};\alpha,\sigma,{\cal H})}
	\approx
	\frac{N^{s/2}\exp[-N S(\pi^\sigma_{\cal H}(\mu)\|\pi^\sigma_{\cal G}(\mu))]}{\alpha^{s/2}\exp[-\alpha S(\pi^\sigma_{\cal H}(\mu)\|\pi^\sigma_{\cal G}(\mu))]}
	,
\label{ratio}
\end{equation}
where $s:=(\dim{\cal H}-\dim{\cal G})$ denotes the number of additional model parameters introduced in the expansion ${\cal G}\to {\cal H}$.
The power factors $N^{s/2}$ and $\alpha^{s/2}$ stem from the normalisation factors of likelihood and prior, respectively,
which depend on the dimension of the theoretical level of description. 

{Bayesian model selection} is thus driven by two main factors \cite{sivia:modelselection,mackay:interpolation}:
(i)
a ratio of exponentials (of which, in the regime $N\gg\alpha$, the denominator can often be approximated by $1$)
favoring the finer-grained model with better fit;
and
(ii)
the ``Occam factor'' $(N/\alpha)^{s/2}$, which favors the simpler model.
It is the trade-off between the exponentials on the one hand,
and the Occam factor on the other, which typically determines whether or not the level of description should be expanded.
If their product is much larger than $1$, one  better stays with the original, coarser-grained description.
In contrast, if it is much less than $1$, one is advised to switch to the finer-grained description.
And if it is of the order $1$, the analysis remains inconclusive, and more data must be collected.

It is important to note that the trade-off decision is not based on experimental data alone.
Rather, it depends also on the initial bias $\sigma$ and on the parameter $\alpha$.
The initial bias constitutes one's starting hypothesis for the model, prior to performing any measurements, and is usually based entirely on symmetry and other theoretical considerations;
whereas $\alpha$ quantifies the associated degree of confidence.
Both $\sigma$ and $\alpha$ reflect  prior expectations of the agent who conducts the experiment, and
so in principle,  carry aspects which remain irreducibly subjective.
In practice, however, rational agents typically agree on the symmetries of the system under study, and hence on a unique initial bias to mirror these symmetries.
In fact, in many cases the initial bias is just equidistribution, $\sigma=1/d$, being maximally non-committal in the absence of any empirical data.
The parameter $\alpha$, on the other hand, can often be estimated \textit{a posteriori} with the help of the evidence procedure.

For large $N$, the estimate for $\alpha$ becomes independent of $N$, and hence the asymptotic behavior of the ratio (\ref{ratio}) is governed entirely by its numerator.
Models can then be selected according to the simple rule of thumb
\begin{equation}
	\chi^2(\pi^\sigma_{\cal H}(\mu)\|\pi^\sigma_{\cal G}(\mu)) / s
	\left\{
\begin{array}{ll}
\ll \ln N & :\,\mbox{keep } {\cal G} \\
\sim\ln N & :\,\mbox{inconclusive} \\
\gg \ln N & :\,\mbox{expand } {\cal G}\to{\cal H}
\end{array}
\right.
.
\label{ruleofthumb}
\end{equation}
Loosely speaking, whenever the gain in accuracy per additional parameter stays below the threshold $\ln N$, 
one better sticks to the simpler model.
Only when this threshold is exceeded, is one advised to move to the finer-grained model with better fit.
The threshold is higher than the threshold for mere statistical significance;
if $1<\chi^2/s<\ln N$ then the potential accuracy gain is significant, yet a refinement of the model is still not recommended.
 
While Bayesian model selection 
is a useful quantitative tool to guide the search for the proper level of description,
it does not amount to an algorithm leading uniquely to ``{the}'' ideal level of description.
The number of possible levels of description is infinite, and while the above framework may help choose between any two of them, 
it cannot replace the creative act of coming up with suitable candidates \cite{mackay:interpolation}.
This creative part is beyond the realm of pure probability, and must involve additional physical considerations such as the study of symmetries, conservation laws, and time scales.

\section{\label{wolf}Wolf's die}

To warm up for the interesting quantum case, I shall illustrate the use of the above mathematical tools in a famous classical example,
Jaynes' analysis of Wolf's die data \cite{jaynes:concentration}.
Rudolph Wolf (1816--1893), a Swiss astronomer, had performed a number
of random experiments, presumably to check the validity of
statistical theory.
In one of these experiments a die was tossed $N=20,000$ times in a
way that precluded any systematic favoring of any face over any other.
The prior expectation was a perfect die, $\sigma=1/6$.
However, the observed relative frequencies $\{f_i\}$ deviated from this expectation;
their measured values are shown in Table \ref{die_table}.
\begin{table}
\begin{center}
\begin{tabular}{c|c|c}
$i$ &
$f_i$ &
$\Delta_i$
\\
\hline
1 &
0.16230 &
-0.00437
\\ 
2 &
0.17245 &
+0.00578
\\ 
3 &
0.14485 &
-0.02182
\\ 
4 &
0.14205 &
-0.02462
\\ 
5 &
0.18175 &
+0.01508
\\ 
6 &
0.19660 &
+0.02993
\\ 
\end{tabular}
\caption{Wolf's die data: frequency distribution $f$ and
its deviation $\Delta$ from the uniform distribution.
\label{die_table}
}
\end{center}
\end{table}
A quick analysis reveals that $1/\sqrt{N}\sim 0.007$, 
so several deviations $\Delta_i$ are outside the typical range.
More precisely, the observed 
$\chi^2(\mu\|\sigma)\approx 271$ lies in the exponential tail far beyond its expected value.
The probability density for such a large $\chi^2$ is extremely small,
$\mbox{pdf}(271|5)\sim 10^{-56}$,
pointing to the presence of systematic defects of the die.

To reflect the presumed nature of the die's imperfections, one may consider a multitude of different levels of description.
Three specific examples are
(i)
the simplest level of description, ${\cal O}=\mbox{span}\{1\}$, corresponding to a Gibbs manifold $\pi^\sigma_{\cal O}({\cal S})$ that consists of the single state $\sigma$ only,
where one stubbornly sticks to the initial bias;
(ii)
at the opposite extreme, the most accurate level of description ${\cal F}$,
where one denies the existence of any simple explanation for the observed deviations and just introduces as many model parameters as data points;
and
(iii)
an intermediate level of description ${\cal G}$, with  two observables characterising the two most likely imperfections.
These are, according to Jaynes:
\begin{itemize}
\item
a shift of the center of gravity due to the mass of ivory
excavated from the spots, which being proportional to the
number of spots on any side, should make the ``observable''
\begin{equation}
G_1^i:=i-3.5
\end{equation}
have a nonzero average.
Indeed, the measured sample mean is $g_1(\mu)=0.0983\ne 0$; 
and
\item
errors in trying to machine a perfect cube, which will tend
to make one dimension (the last side cut) slightly different
from the other two. It is clear from the data that Wolf's die
gave a lower frequency for the faces (3,4); and therefore that
the (3-4) dimension was greater than the (1-6) or (2-5) ones.
The effect of this is that the ``observable''
\begin{equation}
G_2^i :=\left\{
\begin{array}{rl}
1: & i=1,2,5,6 \\
-2: & i=3,4
\end{array}
\right.
\end{equation}
has a nonzero average.
Indeed, $g_2(\mu)=0.1393\ne 0$.
\end{itemize}
If this intermediate level of description turned out to be the most plausible, 
it would provide a genuine explanation, rather than merely a description, of the observed data.

The sample size is large enough to warrant the use of the rule of thumb (\ref{ruleofthumb}).
Successive refinements ${\cal O}\to{\cal G}\to{\cal F}$ of the level of description  
entail additional model parameters and accuracy gains as summarised in Table \ref{die_comparison}.
\begin{table}
\begin{center}
\begin{tabular}{c|c|c|c}
refinement &
$s$ &
$\chi^2$ &
$\chi^2/s$
\\
\hline
${\cal O}\to{\cal G}$ &
2 &
262 &
131
\\ 
${\cal G}\to{\cal F}$ &
3 &
9 &
3
\\
\hline 
${\cal O}\to{\cal F}$ &
5 &
271 &
54
\\ 
\end{tabular}
\caption{Wolf's die data: number of additional model parameters and accuracy gain associated with expansions of the level of description.
\label{die_comparison}
}
\end{center}
\end{table}
Only the first refinement, ${\cal O}\to{\cal G}$, delivers an accuracy gain per additional model parameter that exceeds the threshold $\ln N\approx 10$.
In contrast, the second refinement ${\cal G}\to{\cal F}$, albeit delivering a further accuracy gain that is statistically significant, does not pass this threshold. 
In case the intermediate level of description ${\cal G}$ was not available,
and hence there was a choice only between the ``trivial'' level of description ${\cal O}$ and the ``perfect fit'' level of description ${\cal F}$,
the latter would be more plausible.
Sticking stubbornly to the initial bias is the least plausible of the three options.

If presented with the choice between the three levels of description outlined above, therefore, 
statistical analysis reveals the intermediate, ``explanatory'' level of description to be the most plausible.
This is not to say, however, that this is indeed the best level of description:
One might come up with many more alternative proposals, which would all have to be compared with each other.
Moreover, even if the  above intermediate level of description were confirmed as the winner, statistical analysis would only yield its relative degree of plausibility,
and would never provide certainty about its being the ``true'' level of description.
Statistical analysis cannot replace the creative act of designing levels of description which, as in the example above, are not only supported by the data but also well motivated physically.

\section{\label{qubits}Ising vs. Heisenberg}

Conceptually, Bayesian model selection for quantum systems proceeds in the same way as in the classical case.
The quantumness of the problem enters through the different geometry of the Gibbs manifold. 
As the simplest example, I shall study an exchangeable assembly of qubits;
there, the  geometry to consider is that of the Bloch sphere.

Initially, nothing is known about the qubits, so the prior bias $\sigma$ is uniform.
Then measurements on a sample of $N$ qubits reveal an average Bloch vector of length $r$, 
with an orientation $\hat{n}$ that is tilted by a small angle $\delta\theta$ against the $z$ axis.
The Bloch vector length $r$ is considerably larger than zero, so a new model  is called for which is different from the uniform initial bias.
There might be good physical reasons to expect that the system under consideration is strongly anisotropic in the $z$ direction,
suggesting a level of description ${\cal I}$ (for ``Ising'') comprising the $z$ component of Pauli spin only, ${\cal I}=\mbox{span}\{1,\sigma_z\}$.
In view of the observed tilting angle, however, there might be controversy about this, and a rival proposal (``Heisenberg'') might claim that the level of description should rather include the full Pauli vector, ${\cal H}=\mbox{span}\{1,\sigma_x,\sigma_y,\sigma_z\}$.

To weigh these alternatives in the light of the data, one must evaluate 
\begin{equation}
	\chi^2(\pi^\sigma_{\cal H}(\mu)\|\pi^\sigma_{\cal I}(\mu))/s
	=
	N C^{-1}_{\theta\theta} \delta\theta^2/2
	,
\end{equation}
where $C^{-1}_{\theta\theta}$ denotes the polar component of the entropy-induced metric tensor (\ref{bloch_metric}) on the Bloch sphere.
For instance, for $N=20,000$, $r=0.73$ and a tilting angle of $1$ degree, $\delta\theta=2\pi/360$,
it is
$C^{-1}_{\theta\theta}\approx 0.678$
and $\chi^2/s\approx 2.1$.
Despite a gain in accuracy which is significant, this does not exceed the threshold $\ln N\approx 10$,
and hence Bayesian model selection 
favors the simpler anisotropic model.
At an angle of $2$ degrees, $\chi^2/s$ grows to approximately $8.3$; 
and this being close to the threshold, the analysis remains largely inconclusive.
Finally, for a tilting angle of $3$ degrees, the accuracy gain per additional model parameter attains a value well beyond the threshold, $\chi^2/s\approx 18.6$, 
tipping the balance in favor of the more detailed level of description.

Had one measured a Bloch vector length $r=0.995$ instead of $0.73$, the balance would have tipped in favor of the expanded level of description already at a critical angle of $1$ degree, rather than $2$ degrees.
In general, the more the measured state approaches purity, the more sensitive the choice of level of description becomes to minor directional aberrations from the preferred axis.

\section{\label{conclusions}Conclusions}

Outside the macroscopic domain,
estimating a Gibbs state is a nontrivial inference task,
due to two complicating factors.
First,
for lack of a clear hierarchy of time scales,
the proper set of relevant observables might not be evident \textit{a priori}
but subject to statistical inference.
Second,
whenever experimental data are gathered from a small sample only,
the best estimate for the Lagrange parameters is invariably affected by the experimenter's prior bias.
Both issues can be tackled with the help of Bayesian techniques,
suitably adapted to the problem at hand:
Bayesian model selection, Bayesian interpolation, and the evidence procedure.

The results presented in this paper may have ramifications in a variety of areas.
For the study of thermal properties of a microscopic system
(e.g., a tiny probe taken from a larger system that is presumed to be thermal, or the debris from a single collision experiment)
the framework presented here allows one to decide rationally between rival theories about the proper set of relevant observables,
and subsequently, to find the best estimate for the associated Lagrange parameters.
For incomplete quantum-state tomography, the results imply Bayesian corrections to the conventional maximum entropy scheme;
these corrections become important whenever sample sizes are small.
Moreover, the approach presented here yields not just estimates for the Lagrange parameters, but also the attached error bars.
Finally,
on a conceptual level,
the framework allows for a careful consideration of the thermodynamic limit,
and so may 
shed new light on the long-standing debate about the generality of, or  possible limitations of, the maximum entropy paradigm in statistical mechanics \cite{shore+johnson,tikochinsky:consistent,skilling:axioms,uffink:maxent,uffink:constraintrule}.

I see three avenues for further research.
First,
it will be interesting to see how the Bayesian corrections to conventional state reconstruction schemes play out in practice.
A simple example has been discussed (in the context of the evidence procedure) in Ref. \cite{rau:evidence};
more examples and application to real-world experimental data will be the subject of further work.
Second,
while the model selection framework used here allows  one to assess different proposals for the set of relevant observables,
it does not provide a direct route to the optimal such set.
Doing so requires an extension of  Bayesian reasoning from the space of states to the space of levels of description,
which will be tackled in future work.
Finally,
I consider it worthwhile to study in more detail the asymptotic behavior of the schemes presented here,
in an effort to understand better the emergence of orthodox theory in the macroscopic limit.

\appendix

\section{\label{lod}Level of description}

Any real linear combination of observables is again an observable.
The observables of a physical system thus constitute a real vector space.
This vector space can be endowed with a positive definite scalar product,
the \textit{canonical correlation function} with respect to some \textit{reference state} $\sigma$,
\begin{equation}
	\langle X;Y\rangle_\sigma:=\int_0^1 d\nu\,\mbox{tr}(\sigma^\nu X \sigma^{1-\nu} Y)
	;
\end{equation}
so it is in fact a Hilbert space.
Within this real Hilbert space of observables,
the (typically small) set of observables $\{G_a\}$ which are deemed relevant for the problem at hand,
together with the unit operator, span a proper subspace
\begin{equation}
	{\cal G}:=\mbox{span}\{1,G_a\}
	.
\end{equation}
This subspace is termed the \textit{level of description} \cite{rau:physrep}.

Levels of description might be related by coarse graining or complementation.
A level of description ${\cal G}$ is ``coarser'' than another level of description ${\cal F}$, ${\cal G}\subset {\cal F}$, if the former is a subspace of the latter.
The coarse graining relation $\subset$ induces a partial ordering of the levels of description,
with unique minimal element ${\cal O}:=\mbox{span}\{1\}$ and maximal element ${\cal A}$, the total Hilbert space of observables.
The level of description ${\cal G}$ is ``complementary'' to ${\cal F}$,
${\cal G}=\neg_{{\cal A},\sigma}{\cal F}$,
if observables from both levels of description together span the entire space of observables, 
and if in the reference state $\sigma$ the two levels are uncorrelated,
\begin{eqnarray}
	{\cal G}=\neg_{{\cal A},\sigma}{\cal F}
	&:\Leftrightarrow&
	\mbox{span}\{1,G_a,F_b\}={\cal A}
	\ ,\ 
	\nonumber \\
	&&
	\langle \delta X;\delta Y\rangle_\sigma=0
	\ \forall X\in{\cal G}, Y\in{\cal F}
	,
\end{eqnarray}
with $\delta X:=X-\langle X\rangle_\sigma$.
Complementation reverses the direction of coarse graining,
\begin{equation}
	{\cal G}\subset {\cal F} \Rightarrow \neg_{{\cal A},\sigma}{\cal F} \subset \neg_{{\cal A},\sigma}{\cal G}
	;
\end{equation}
and when applied twice, it returns the original level of description,
\begin{equation}
	\neg_{{\cal A},\sigma}\neg_{{\cal A},\sigma}{\cal G} = {\cal G}
	.
\end{equation}
The properties of
coarse graining and complementation are reminiscient of those of logical implication and negation.
In this sense, one may say that the space of observables gives rise to a minimal logical structure.

The intersection and closed hull of two levels of description are denoted by ${\cal G}\cap {\cal F}$ and ${\cal G}\cup {\cal F}$, respectively.
In line with the logical structure mentioned above, the operations $\cap,\cup$ share some properties with the Boolean ``and'' and ``or'' operations such as commutativity, associativity and reversal under complementation;
yet the analogy is not perfect since in contrast to classical Boolean logic, they violate distributivity.
If the levels of description pertain to two different physical systems $A$ and $B$ then it is
${\cal G}^A\cap {\cal F}^B={\cal O}^{AB}$, and
\begin{equation}
	{\cal G}^A\cup {\cal F}^B
	=
	\mbox{span}\{1^A\otimes 1^B, G_a^A\otimes 1^B, 1^A\otimes F_b^B\}
	.
\end{equation}
A further way to concatenate the two constituent levels of description is by means of the tensor product
\begin{equation}
	{\cal G}^A\otimes {\cal F}^B
	:=
	\mbox{span}\{1^A\otimes 1^B, G_a^A\otimes 1^B, 1^A\otimes F_b^B, G_a^A\otimes F_b^B\}
	.
\end{equation}

\section{\label{relpart}Relevant part of a state}

For an arbitrary state $\rho$, its \textit{relevant part} with respect to a level of description ${\cal G}$ and reference state $\sigma$ is the unique state $\pi^\sigma_{\cal G}(\rho)$ 
which for all observables in the level of description yields the same expectation values as $\rho$,
yet within this constraint,
is as close as possible to the reference state.
The distance to the reference state is measured in terms of the
\textit{relative entropy} \cite{ochs:properties,wehrl:rmp,donald:cmp,vedral:rmp}
\begin{equation}
	S(\rho\|\sigma):= \mbox{tr}(\rho\ln\rho-\rho\ln\sigma)
	.
\end{equation}
The relevant part  is thus determined by the minimization
\begin{equation}
	S(\pi^\sigma_{\cal G}(\rho)\|\sigma)=\min_{g(\rho')=g(\rho)} S(\rho'\|\sigma)
	,
\end{equation}
where I employed $g(\rho')$ as a shorthand notation for the set $\{\langle G_a\rangle_{\rho'}\}$.

That the relative entropy is the appropriate measure for the distance between two states,
follows from the \textit{quantum Stein lemma} \cite{hiai+petz,ogawa+nagaoka,brandao+plenio}.
According to this lemma,
given a finite sample of size $N$ taken from an i.i.d. source of states $\sigma$,
the probability that
tomography on this sample will erroneously reveal some different state $\rho$,
\begin{eqnarray}
	&&{
	\mbox{prob}_{1-\epsilon}(\rho|N,\sigma):=
	}
	\nonumber \\
	&&
	\inf_{\Gamma} \left.\left\{\mbox{prob}(\Gamma|\sigma^{\otimes N})\right|\mbox{prob}(\Gamma|\rho^{\otimes N})\geq 1-\epsilon\right\}
	,
\label{defprob}
\end{eqnarray}
decreases asymptotically as 
\begin{equation}
	\mbox{prob}_{1-\epsilon}(\rho|N,\sigma)\sim \exp[-N S(\rho\|\sigma)]
	,
\label{quantumstein}
\end{equation}
regardless of the specific value of the error parameter  $\epsilon$ ($0<\epsilon<1$).
The $\Gamma$ featuring in the above definition are propositions (projection operators) about the sample 
which asymptotically, i.e., to within an error probability $\epsilon$ that does not depend on sample size, are compatible with the sample being in the state $\rho^{\otimes N}$.
Taking the infimum over $\Gamma$ picks that proposition which is most confined, and hence  discriminates best between $\sigma$ and $\rho$.
The coefficient in the exponent is the relative entropy between the two states, which is thus recognised as 
the proper measure of their distinguishability  \cite{vedral:rmp}.

The relevant part of a state has the generalised Gibbs form \cite{ruskai:minrent}
\begin{equation}
	\pi^\sigma_{\cal G}(\rho) 
	= 
	Z(\lambda)^{-1} \exp\left[(\ln\sigma - \langle\ln\sigma\rangle_\sigma)-\sum_{a=1}^r \lambda^a G_a\right]
	,
\label{canonical}
\end{equation}
with the \textit{partition function}  
\begin{equation}
	Z(\lambda):=\mbox{tr}\left\{\exp\left[(\ln\sigma - \langle\ln\sigma\rangle_\sigma)-\sum_{a=1}^r \lambda^a G_a\right]\right\}
\end{equation}
ensuring state normalisation,
and the \textit{Lagrange parameters} $\{\lambda^a\}$ adjusted such that 
$g(\pi^\sigma_{\cal G}(\rho))=g(\rho)$.
Amongst all states of the above generalised Gibbs form,
the relevant part of $\rho$ is that which comes closest to $\rho$ in terms of relative entropy,
\begin{equation}
	S(\rho\|\pi_{\cal G}^\sigma(\rho))
	=
	\min_{\rho'} S(\rho\|\pi_{\cal G}^\sigma(\rho'))
	.
\end{equation}
The reference state is often, but not always, the uniform distribution;
if so, the generalised Gibbs state acquires the more familiar form
\begin{equation}
	\pi_{\cal G}(\rho) 
	= 
	Z(\lambda)^{-1} \exp\left[-\sum_{a=1}^r \lambda^a G_a\right]
\end{equation}
(with  superscript $\sigma$ omitted) 
which maximizes the  von Neumann entropy $S[\rho]:=-\mbox{tr}(\rho\ln\rho)$
under the given constraints.

Since the relevant part of a state retains complete information solely about selected degrees of freedom 
(the observables contained in the {level of description}),
while discarding information about the rest,
the map $\pi_{\cal G}^\sigma:\rho\to\pi^\sigma_{\cal G}(\rho)$ may be regarded as a \textit{coarse graining} operation.
Indeed, this operation bears some resemblance to a projection operator:
it is idempotent,
\begin{equation}
	\pi^\tau_{\cal G}\circ\pi^\sigma_{\cal G} = \pi^\tau_{\cal G}
\end{equation}
(even for $\tau\neq\sigma$);
successive coarse grainings with smaller and smaller levels of description are equivalent to a one-step coarse graining with the smallest level of description,
\begin{equation}
	{\cal G} \subset {\cal F} 
	\ \Leftrightarrow \ 
	\pi_{\cal G}^\sigma\circ\pi_{\cal F}^\sigma = \pi_{\cal G}^\sigma
	;
\end{equation}
and it is covariant under unitary transformations,
\begin{equation}
	\pi^{U\sigma U^\dagger}_{U{\cal G}U^\dagger}(U\rho U^\dagger) = U \pi^\sigma_{\cal G}(\rho) U^\dagger
	.
\end{equation}
In contrast to a true projection operator, however, the coarse graining map is in general not linear.
In case of a uniform reference state, the coarse graining map is the (possibly nonlinear) dual of the Kawasaki-Gunton projector, a projection superoperator acting on the space of observables \cite{kawasaki+gunton}.

\section{\label{geometry}Gibbs manifold}

Let ${\cal S}$ denote the set of all normalised mixed states of a given physical system.
This set constitutes a differentiable manifold of dimension $(d^2-1)$, where $d$ is the Hilbert space dimension.
In this manifold, states of the generalised Gibbs form (\ref{canonical}) constitute a submanifold $\pi_{\cal G}^\sigma({\cal S})$;
I call it the \textit{Gibbs manifold} associated with level of description ${\cal G}$ and reference state $\sigma$.
A point on this Gibbs manifold, and hence a specific state of generalised Gibbs form, is a \textit{Gibbs model}.
The Gibbs manifold has dimension
\begin{equation}
	\dim \pi_{\cal G}^\sigma({\cal S}) = \dim {\cal G} -1
	,
\end{equation}
which equals the number of relevant observables $\{G_a\}$
as long as these are linearly independent.
Coordinates on the manifold may be the Lagrange parameters $\{\lambda^a\}$ or the expectation values $\{g_a\}$,
or any set of $(\dim {\cal G} -1)$ independent functions thereof.
Lagrange parameter coordinates are related to expectation value coordinates via 
\begin{equation}
	g_a=-{\partial} (\ln Z) / {\partial \lambda^a}
	.
\end{equation}

Upon infinitesimal variation of the Lagrange parameters, the expectation value of an arbitrary observable $A$
changes by
\begin{equation}
	d\langle A\rangle
	=
	-\sum_a \langle \delta G_a; A\rangle d\lambda^a
	,
\end{equation}
with $\delta G_a:= G_a-g_a$,
and the expectation values and the canonical correlation function evaluated in the model with coordinates $\{\lambda^a\}$.
A special case is the variation of relevant expecation values,
\begin{equation}
	d g_b = -\sum_a d\lambda^a C_{ab}
	,
\end{equation}
where the coefficients
\begin{equation}
	C_{ab}:=\langle \delta G_a;\delta G_b\rangle
	=
	\frac{\partial^2}{\partial\lambda^a\partial\lambda^b}\ln Z
\end{equation}
form the $r\times r$ \textit{correlation matrix}.
As the canonical correlation function has all properties of a positive definite scalar product in the  space of observables, the correlation matrix is symmetric and positive.

The Gibbs manifold is endowed with a natural Riemannian metric and volume element, induced by the relative entropy \cite{balian:physrep,skilling:classic}.
As one would expect from a proper distance measure,
the relative entropy between two states is always positive,
\begin{equation}
	S(\rho\|\rho')\geq 0
	,
\end{equation}
with equality if and only if $\rho=\rho'$;
and even though it is in general not symmetric, $S(\rho\|\rho')\neq S(\rho'\|\rho)$,
it is approximately so for nearby states:
\begin{equation}
	S(\rho\|\rho+\delta\rho)\sim O((\delta\rho)^2)
	.
\end{equation}
The relative entropy between two points $\omega,\omega'$ on the same Gibbs manifold is
\begin{equation}
	S(\omega\|\omega')
	=
	\sum_a ({\lambda'}^a -\lambda^a) g_a + (\ln{Z'}-\ln Z)
	;
\end{equation}
which for nearby states is approximately quadratic in the coordinate differentials, 
\begin{eqnarray}
	S(\omega\|\omega+\delta\omega)
	&\approx&
	(1/2) \sum_{ab} C_{ab} \delta \lambda^a\delta\lambda^b
	\nonumber \\
	&\approx&
	(1/2) \sum_{ab} (C^{-1})^{ab} \delta g_a \delta g_b
	.
\label{relentropy_nearby}
\end{eqnarray}
The {correlation matrix} $C$ or its inverse $C^{-1}$, respectively, may thus be regarded as 
a {metric tensor} on the Gibbs manifold.
Associated with this  metric is the volume element
\begin{equation}
	\int_{\pi^\sigma_{\cal G}({\cal S})} d\omega
	=
	\int \prod_a d\lambda^a \sqrt{\det C}
	=
	\int \prod_a d g_a \sqrt{\det C^{-1}}	
	.
\label{volume}
\end{equation}

Given some coarser level of description ${\cal H}$, ${\cal H}\subset {\cal G}$,
the Gibbs manifold $\pi^\sigma_{\cal G}({\cal S})$ can be viewed 
as a fiber bundle,
with the reduced Gibbs manifold $\pi^\sigma_{\cal H}({\cal S})$ as its base, and the coarse graining map $\pi_{\cal H}^\sigma$
as the bundle projection,
\begin{equation}
	\pi_{\cal H}^\sigma : \pi^\sigma_{\cal G}({\cal S}) \ni\omega \to \zeta \in \pi^\sigma_{\cal H}({\cal S})
	.
\end{equation}
The fiber over $\zeta$ is the submanifold of Gibbs models satisfying the constraint
$h(\omega) = h(\zeta)$,
\begin{equation}
	\pi^\sigma_{\cal G} \circ (\pi_{\cal H}^\sigma)^{-1}(\zeta) = \pi^\sigma_{\cal G}({\cal S})|_{h(\zeta)}
	.
\end{equation}
It is then possible to factorize volume elements of the original Gibbs manifold into those of its fiber and base,
\begin{equation}
	\int_{\pi^\sigma_{\cal G}({\cal S})} d\omega
	=
	\int_{\pi^\sigma_{\cal H}({\cal S})} d\zeta \int_{{\pi^\sigma_{\cal G}({\cal S})}|_{h(\zeta)}} d\omega
	.
\label{factorize}
\end{equation}

\section{\label{bloch}Geometry of the Bloch sphere}

Any normalised mixed state of a single qubit can be written as 
\begin{equation}
	\rho= (1/2) (1+\langle\vec{\sigma}\rangle_\rho \cdot\vec{\sigma})
	,
\end{equation}
with $\vec{\sigma}$ defined as the vector of Pauli matrices, $\vec{\sigma}:=(\sigma_x,\sigma_y,\sigma_z)$.
The expectation value of the latter is the \textit{Bloch vector;}
it has the spatial direction $\hat{n}$ and length $r$, 
\begin{equation}
	\langle\vec{\sigma}\rangle_\rho = r \hat{n}
	.
\end{equation}
The Pauli matrices being  informationally complete,
the above state can always be brought into the Gibbs form
\begin{equation}
	\rho=Z(\vec{\lambda})^{-1} \exp(-\vec{\lambda}\cdot\vec{\sigma})
	,
\end{equation}
with 
Lagrange parameters
\begin{equation}
	\vec{\lambda}=-(\tanh^{-1}r) \hat{n}
\end{equation}
and partition function
\begin{equation}
	Z(\vec{\lambda})=2\cosh|\vec{\lambda}|=2/\sqrt{1-r^2}
	.
\end{equation}

The relative entropy between two arbitrary qubit states is
\begin{eqnarray}
	S(\rho\|\rho')
	&=&
	r \tanh^{-1} r - r \tanh^{-1} r' (\hat{n}\cdot \hat{n}')
	+
	\nonumber \\
	&&
	(1/2) \ln(({1-r^2})/({1-r'^2}))
	;
\end{eqnarray}
which for nearby states becomes approximately
\begin{equation}
	S(\rho\|\rho')
	\approx
	(1/2) [C^{-1}_{rr} \delta r^2 + C^{-1}_{\theta\theta} \delta\theta^2 + C^{-1}_{\phi\phi} \delta\phi^2]
	.
\end{equation}
Here $(r,\theta,\phi)$ are the spherical coordinates of the Bloch vector as defined by
\begin{equation}
	\langle\sigma_x\rangle=r\sin\theta\cos\phi
	\ ,\ 
	\langle\sigma_y\rangle=r\sin\theta\sin\phi
	\ ,\ 
	\langle\sigma_z\rangle=r\cos\theta
	,
\end{equation}
with $r\in[0,1]$, $\theta\in[0,\pi]$ and $\phi\in[0,2\pi)$;
and
$C^{-1}$ denotes the entropy-induced metric tensor (inverse of the correlation matrix) on the Bloch sphere.
In spherical coordinates this metric tensor is diagonal,
\begin{equation}
	C^{-1}=\mbox{diag}(1/(1-r^2), r\tanh^{-1}r, r\tanh^{-1}r \sin^2\theta)
	,
\label{bloch_metric}
\end{equation}
but differs from the ordinary  
metric $\mbox{diag}(1,r^2,r^2\sin^2\theta)$.
Consequently, the associated volume element
\begin{equation}
	\sqrt{\det C^{-1}}
	=
	r\tanh^{-1}r \sin\theta / \sqrt{1-r^2}
	,
\end{equation}
too, differs from its ordinary counterpart, especially near the surface of the Bloch sphere:
\begin{equation}
	\frac{\sqrt{\det C^{-1}}}{r^2 \sin\theta}
	=
	\frac{\tanh^{-1}r}{r\sqrt{1-r^2}}
	\left\{
\begin{array}{ll}
\approx 1 & :\,r\ll 1 \\
\to\infty & :\,r\to 1
\end{array}
\right.
.
\end{equation}
Distinguishable quantum states are thus not spread uniformly throughout the Bloch sphere as one might expect classically, 
but are concentrated on or near its surface.

\section{\label{entropic}Entropic distribution}

The coordinates of a Gibbs model $\omega\in\pi^\sigma_{\cal G}({\cal S})$, and hence its location on the Gibbs manifold, might not be precisely known
but have some probability distribution.
Such a distribution over the Gibbs manifold is \textit{entropic}, $\omega\sim\mbox{Ent}(\alpha,\sigma,{\cal G})$, 
if it has the form
\begin{equation}
	\mbox{prob}(\omega|\alpha,\sigma,{\cal G})
	\propto
	\left\{
\begin{array}{ll}
\exp[-\alpha S(\omega\|\sigma)] & :\,\omega\in\pi^\sigma_{\cal G}({\cal S}) \\
0 & :\,\mbox{else}
\end{array}
\right.
	,
\label{exponential_ansatz}
\end{equation}
with $\alpha>0$ and a factor of proportionality that does not depend on $\omega$.
For large $\alpha$ this is approximately a Gaussian on $\pi^\sigma_{\cal G}({\cal S})$ of width $1/\sqrt{\alpha}$ around the reference state $\sigma$. 

The entropic distribution has a number of important properties.
(i) 
If $\omega$ is entropically distributed then so is $U\omega U^\dagger$ for any unitary $U$, with co-transformed reference state and level of description,
\begin{equation}
	\mbox{prob}(U\omega U^\dagger|\alpha,U\sigma U^\dagger,U{\cal G}U^\dagger)
	=
	\mbox{prob}(\omega|\alpha,\sigma,{\cal G})
	.
\end{equation}
(ii)
Coarse graining ${\cal G}\to{\cal H}\subset{\cal G}$ leaves relative probabilities invariant,
\begin{equation}
	\mbox{prob}(\pi^\sigma_{\cal H}(\omega)|\alpha,\sigma,{\cal H})
	\propto
	\mbox{prob}(\pi^\sigma_{\cal H}(\omega)|\alpha,\sigma,{\cal G})
	,
\label{extension_prior}
\end{equation}
with a factor of proportionality which is independent of $\omega$.
(iii)
If the reference state is uncorrelated then 
the entropic distribution does not introduce any bias towards spurious correlations,
\begin{eqnarray}
	&&
	\mbox{prob}(\omega^{AB}|\alpha,{\sigma^A\otimes\sigma^B},{\cal G}^A\otimes {\cal F}^B)
	\leq
	\nonumber \\
	&&
	\mbox{prob}(\omega^{A}\otimes\omega^{B}|\alpha,{\sigma^A\otimes\sigma^B},{\cal G}^A\otimes {\cal F}^B)
	,
\end{eqnarray}
where $\omega^{A},\omega^B$ are the respective reductions of $\omega^{AB}$.
And
(iv)
for uncorrelated states, the probability factorises,
\begin{eqnarray}
	&&
	\mbox{prob}(\omega^{A}\otimes\omega^{B}|\alpha,{\sigma^A\otimes\sigma^B},{\cal G}^A\otimes {\cal F}^B)
	\propto
	\nonumber \\
	&&
	\mbox{prob}(\omega^{A}|\alpha,{\sigma^A},{\cal G}^A)\,
	\mbox{prob}(\omega^{B}|\alpha,{\sigma^B},{\cal F}^B)
	.
\label{prior_factorise}
\end{eqnarray}
If the reference state $\sigma$ is uniform then 
the entropic distribution is in fact the \textit{only} probability distribution with the above four properties,
both in the classical \cite{aczel:natural} and in the quantum case \cite{ochs:axiomatic}.
In contrast, for arbitrary $\sigma$ the uniqueness of the entropic distribution has been shown in the classical case only \cite{ochs:unique};
but I conjecture that this result, too, should carry over to the quantum case.

Finally, the product of two entropic distributions is again entropic,
\begin{eqnarray}
	&&
	\mbox{prob}(\omega|N,\mu,{\cal G}) \, \mbox{prob}(\omega|\alpha,\sigma,{\cal G})
	\propto
	\nonumber \\
	&&
	\mbox{prob}(\omega|\alpha+N,\rho(\mu,\sigma;t),{\cal G})
	,
\end{eqnarray}
provided they are defined over the same Gibbs manifold, $\pi^\mu_{\cal G}({\cal S})=\pi^\sigma_{\cal G}({\cal S})$.
Here $\rho(\mu,\sigma;t)\in\pi^\sigma_{\cal G}({\cal S})$ denotes the interpolated reference state
\begin{equation}
	\rho(\mu,\sigma;t)
	\propto
	\exp[(1-t)\ln\mu + t\ln\sigma]
\label{interpolated}
\end{equation}
with $t:=\alpha/(\alpha+N)$.

\section{\label{approximations}Gaussian approximation and thermodynamic limit}

In many practical applications the states under consideration are all concentrated inside some small region of state space.
It is then often justified to make the \textit{Gaussian approximation}, in which relative entropies are quadratic in the coordinate differentials. 
In this approximation 
the relative entropy is symmetric,
\begin{equation}
	S(\rho\|\omega)\approx S(\omega\|\rho)
	.
\end{equation}
Furthermore, the normalisation factor of the entropic distribution becomes 
\begin{equation}
	\int_{\pi^\sigma_{\cal G}({\cal S})} d\omega \,
	\exp[-\alpha S(\omega\|\sigma)]
	\approx
	(2\pi/\alpha)^{\dim\pi^\sigma_{\cal G}({\cal S})/2}
	,
\end{equation}
and thus independent not only of $\omega$ but also of $\sigma$.
As a consequence, the entropic distribution becomes invariant under exchange of $\omega$ and $\sigma$,
\begin{equation}
	\mbox{prob}(\omega|\alpha,\sigma,{\cal G})
	\approx
	\mbox{prob}(\sigma|\alpha,\omega,{\cal G})
	.
\end{equation}

When the Gibbs manifold $\pi^\sigma_{\cal G}({\cal S})$ is considered as a fiber bundle (see Appendix \ref{geometry}),
with some reduced manifold $\pi^\sigma_{\cal H}({\cal S})$, 
${\cal H}\subset{\cal G}$,
as its base,
then in the Gaussian approximation the fiber over $\zeta\in\pi^\sigma_{\cal H}({\cal S})$ is given by
\begin{equation}
	\pi^\sigma_{\cal G} \circ (\pi_{\cal H}^\sigma)^{-1}(\zeta) 
	\approx 
	\pi^\zeta_{\neg_{{\cal G},\zeta}{\cal H}}({\cal S})
	.
\end{equation}
Moreover, for any $\omega\in\pi^\sigma_{\cal G}({\cal S})$ the four states
\begin{equation}
	\begin{array}{ccc}
		\pi^\sigma_{\cal H}(\omega) & \mbox{------} & \omega \\
		| & & | \\
		\sigma & \mbox{------} & \pi^\sigma_{\neg_{{\cal G},\sigma}{\cal H}}(\omega)
	\end{array}
\end{equation}
form  a rectangle as shown,
with opposite sides having approximately equal length as measured by the relative entropy.
Together with the (exact)
\textit{law of Pythagoras} \cite{petz:book} for the relative entropy,
\begin{equation}
	S(\omega\|\sigma)
	=
	S(\omega\|\pi_{\cal H}^\sigma(\omega)) + S(\pi_{\cal H}^\sigma(\omega)\|\sigma)
	,
\label{pythagoras}
\end{equation}
these properties imply that the entropic distribution factorises into separate distributions over fiber and base,
\begin{eqnarray}
	\mbox{prob}(\omega|\alpha,\sigma,{\cal G})
	&\approx&
	\mbox{prob}(\pi^\sigma_{\neg_{{\cal G},\sigma}{\cal H}}(\omega)|\alpha,\sigma,\neg_{{\cal G},\sigma}{\cal H})
	\times
	\nonumber \\
	&&
	\mbox{prob}(\pi^\sigma_{\cal H}(\omega)|\alpha,\sigma,{\cal H})
	.
\end{eqnarray}

If a model $\mu$ is entropically distributed,
$\mu\sim\mbox{Ent}(N,\omega,{\cal F})$,
around a reference state $\omega$ which  is itself entropically distributed,
$\omega\sim\mbox{Ent}(\alpha,\sigma,{\cal G})$,
then in the Gaussian approximation and for ${\cal F}\supset{\cal G}$, the product of these entropic distributions is
\begin{eqnarray}
	&&
	\mbox{prob}(\mu|N,\omega,{\cal F})\,
	\mbox{prob}(\omega|\alpha,\sigma,{\cal G})
	\propto
	\mbox{prob}(\mu|N,\rho ,{\cal F})
	\times
	\nonumber \\
	&&
	\mbox{prob}(\omega|\alpha+N,\rho ,{\cal G})
	\,
	\mbox{prob}(\rho |\alpha,\sigma,{{\cal F}\cap{\cal G}})
	,
\label{post_supset}
\end{eqnarray}
with a factor of proportionality that is independent of both $\mu$ and $\omega$,
and with $\rho \in\pi^\sigma_{{\cal F}\cap{\cal G}}({\cal S})$
short for the interpolated state 
$\rho(\pi^\sigma_{{\cal F}\cap{\cal G}}(\mu),\sigma;t)$ as defined in Eq. (\ref{interpolated}).
In the case ${\cal F}\subset{\cal G}$, 
the product is approximately
\begin{eqnarray}
	&&
	\mbox{prob}(\mu|N,\omega,{\cal F})\,
	\mbox{prob}(\omega|\alpha,\sigma,{\cal G})
	\propto
	\nonumber \\
	&&
	\mbox{prob}(\pi^{\sigma}_{\cal F}(\mu)|N,\rho ,{\cal F})
	\,
	\mbox{prob}(\pi^{\sigma}_{\cal F}(\omega)|\alpha+N,\rho ,{\cal F})
	\times
	\nonumber \\
	&&
	\mbox{prob}(\pi^{\rho}_{\neg_{{\cal G},\rho}{\cal F}}(\omega)|\alpha,\rho,{\neg_{{\cal G},\rho}{\cal F}})
	\,
	\mbox{prob}(\rho |\alpha,\sigma,{{\cal F}\cap{\cal G}})
	.
\label{post_subset}
\end{eqnarray}

In the \textit{thermodynamic limit} the parameter $N$ (but not the parameter $\alpha$)
approaches infinity, $N\to\infty$.
The interpolation parameter $t$ then approaches zero, $t\to 0$,
and as a consequence, $\rho\to\pi^\sigma_{{\cal F}\cap{\cal G}}(\mu)$.
In this limit the above product approaches asymptotically, for ${\cal F}\supset{\cal G}$,
\begin{eqnarray}
	&&
	\mbox{prob}(\mu|N,\omega,{\cal F})\,
	\mbox{prob}(\omega|\alpha,\sigma,{\cal G})
	\sim
	\delta_{\neg_{{\cal F},\sigma}{\cal G}}(\mu-\rho)
	\times
	\nonumber \\
	&&
	\delta_{\cal G}(\omega-\rho)
	\,
	\mbox{prob}(\rho|\alpha,\sigma,{{\cal F}\cap{\cal G}})
	,
\end{eqnarray}
where $\delta_{\neg_{{\cal F},\sigma}{\cal G}}$ and $\delta_{\cal G}$ are multi-dimensional delta functions on
the Gibbs manifolds $\pi^\rho_{\neg_{{\cal F},\sigma}{\cal G}}({\cal S})$ and $\pi^\rho_{\cal G}({\cal S})$, respectively;
and for ${\cal F}\subset{\cal G}$,
\begin{eqnarray}
	&&
	\mbox{prob}(\mu|N,\omega,{\cal F})\,
	\mbox{prob}(\omega|\alpha,\sigma,{\cal G})
	\sim
	\delta_{\cal F}(\pi^\sigma_{\cal F}(\omega)-\rho)\times
	\nonumber \\
	&&
	\mbox{prob}(\pi^{\rho}_{\neg_{{\cal G},{\rho}}{\cal F}}(\omega)|\alpha,{\rho},{\neg_{{\cal G},{\rho}}{\cal F}})\,
	\mbox{prob}(\rho|\alpha,\sigma,{{\cal F}\cap{\cal G}})
	.
\end{eqnarray}

\section{\label{statmech}Link to thermodynamics}

Much of conventional thermodynamics amounts to exploring the differential geometry of the Gibbs manifold,
and in particular,
transforming its coordinates to that set of variables which is best suited for the problem at hand. 
Contained in this set is usually the
\textit{thermodynamic entropy},
which for a Gibbs model $\omega\in\pi^\sigma_{\cal G}({\cal S})$ 
is defined as 
\begin{equation}
	S:= -S(\omega\|\sigma) -\langle\ln\sigma\rangle_\sigma
	.
\end{equation}
(I employ natural units with $k_B=1$.)
If the reference state is uniform, this reduces to the more familiar expression 
$S=- \langle\ln\omega\rangle_\omega$.
The thermodynamic entropy is related to the Lagrange parameters and expectation values via 
\begin{equation}
	S=\ln Z +\sum_a \lambda^a g_a
	,
\end{equation}
with differential
\begin{equation}
	dS = \sum_a \lambda^a dg_a
	.
\end{equation}

In addition to the Lagrange parameters $\{\lambda^a\}$, the partition function $Z$ might depend on further parameters $\{\xi^b\}$;
these might be, say, parameters that determine the choice of Hilbert space (e.g., a fixed spatial volume or particle number)
or control parameters on which the operators $\{G_a\}$ depend (e.g., an external field).
Associated with these parameters $\{\xi^b\}$ are then further variables
\begin{equation}
	\kappa_b:= {\partial} (\ln Z)/{\partial\xi^b}
	.
\end{equation}
Taking these into account, the entropy differential reads
\begin{equation}
	dS=\sum_a \lambda^a dg_a + \sum_b \kappa_b d\xi^b
	.
\label{entropydifferential}
\end{equation}

Being the only assured constant of the motion, the \textit{internal energy} $U$  features always as a  variable in conventional thermodynamics,
with the \textit{inverse temperature} $\beta$ as its conjugate.
Depending on whether the energy is given on average or as a sharp constraint,
the pair $(U,\beta)$ may be 
of the type $(g,\lambda)$ (``canonical ensemble'')
or $(\xi,\kappa)$ (``microcanonical ensemble''), respectively.
In both cases, defining the \textit{temperature}
\begin{equation}
	T:=1/\beta
\end{equation}
and new variables 
\begin{equation}
	w^a:= -\lambda^a/\beta
	\ ,\ 
	X_b:= -\kappa_b / \beta
	,
\end{equation}
one  obtains from Eq. (\ref{entropydifferential}) the \textit{first law of thermodynamics},
\begin{equation}
	dU = \underbrace{T dS}_{\delta Q} + \underbrace{\sum_a w^a dg_a + \sum_b X_b d\xi^b}_{\delta W}
	.
\end{equation}
Here the differential $\delta Q$ denotes \textit{heat}, and $\delta W$ denotes \textit{work}.
Some common choices for the pairs $(g,w)$ and $(\xi,X)$ are listed in Table \ref{variables}.

\begin{table}
\begin{center}
\begin{tabular}{c|c|l}
$(g,w)$ &
$(\xi,X)$ &
names
\\
\hline
$(\vec{p},\vec{v})$ &
 &
momentum, velocity
\\ 
$(\vec{L},\vec{\omega})$ &
 &
angular momentum, angular velocity
\\ 
$(N,\mu)$ &
$(N,\mu)$ &
particle number, chemical potential
\\ 
$(\vec{M},\vec{B})$ &
$(\vec{B},-\vec{M})$ &
magnetic field, magnetization
\\ 
$(\vec{P},\vec{E})$ &
$(\vec{E},-\vec{P})$ &
electric field, electric polarization
\\ 
 &
$(V,-p)$ &
volume, pressure
\\ 
\end{tabular}
\caption{Common examples of thermodynamic variables.
In cases where two alternative pairings are given, the proper choice depends on the specific situation:
For instance, $(\vec{M},\vec{B})$ should be used if the magnetization is an (approximate) constant of the motion and given on average,
whereas $(\vec{B},-\vec{M})$ should be employed if the magnetic field is an external control parameter for the Hamiltonian.
\label{variables}
}
\end{center}
\end{table}

The internal energy $U$ is an example of a \textit{thermodynamic potential}.
Other important examples are the \textit{free energy} 
\begin{equation}
	F:= U-TS
\end{equation}
and the \textit{grand potential} 
\begin{equation}
	A:= U-TS -\sum_a w^a g_a
	,
\end{equation}
with respective differentials
\begin{equation}
	dF=-S dT+\sum_a w^a dg_a +\sum_b X_b d\xi^b
\end{equation}
and
\begin{equation}
	dA = -S dT -\sum_a g_a dw^a + \sum_b X_b d\xi^b
	.
\end{equation}
The latter implies, e.g.,
$S=-(\partial A/\partial T)_{w,\xi}$,
where the subscripts denote the variables to be kept fixed when taking the partial derivative.

The grand potential is directly linked to the partition function,
\begin{equation}
	A(T,w^a,\xi^b) = - T \ln Z(T,w^a,\xi^b)
	,
\end{equation}
which in turn can be calculated microscopically.
A key part of statistical mechanics is determining the partition function and hence the grand potential, 
and subsequently relating the latter, via suitable coordinate transformations on the Gibbs manifold, to the other thermodynamic variables of interest.

\begin{acknowledgments}
I thank Gernot Alber and Joe Renes for stimulating discussions.
\end{acknowledgments}


%

\end{document}